**RESEARCH ARTICLE**  **Open Access**

# eXamine: Exploring annotated modules in networks

Kasper Dinkla[1†], Mohammed El-Kebir[2,3†], Cristina-Iulia Bucur[2,3], Marco Siderius[3], Martine J Smit[3], Michel A Westenberg[1*] and Gunnar W Klau[2,3]

**Abstract**

**Background:** Biological networks have a growing importance for the interpretation of high-throughput "omics" data. Integrative network analysis makes use of statistical and combinatorial methods to extract smaller subnetwork modules, and performs enrichment analysis to annotate the modules with ontology terms or other available knowledge. This process results in an annotated module, which retains the original network structure and includes enrichment information as a set system. A major bottleneck is a lack of tools that allow exploring both network structure of extracted modules and its annotations.

**Results:** This paper presents a visual analysis approach that targets small modules with many set-based annotations, and which displays the annotations as contours on top of a node-link diagram. We introduce an extension of self-organizing maps to lay out nodes, links, and contours in a unified way. An implementation of this approach is freely available as the Cytoscape app eXamine

**Conclusions:** eXamine accurately conveys small and annotated modules consisting of several dozens of proteins and annotations. We demonstrate that eXamine facilitates the interpretation of integrative network analysis results in a guided case study. This study has resulted in a novel biological insight regarding the virally-encoded G-protein coupled receptor US28.

## Background

High-throughput "omics" data provide snapshots of cellular states in a specific condition. Computational approaches can be used to relate these low-level measurements with high-level changes in phenotype. Traditionally, these approaches were *gene-centric* and typically resulted in ranked lists of differentially expressed genes [1-3]. Later, gene-centric approaches were complemented by *pathway-* [4,5] and *network-based* methods [6,7] to provide inter-gene context for mechanistic insights. Pathway-based approaches identify overrepresented pathways from databases such as the Kyoto Encyclopedia of Genes and Genomes (KEGG) [8]. Network-based approaches yield small, *de novo* subnetwork modules that may span several known pathways, and reveal their crosstalk [9].

Extracted network modules are analyzed in the context of established gene annotations to hypothesize about the module's role in high-level cell conditions (see Figure 1). Genes are often related to very many terms (too many for human comprehension), most of which are likely irrelevant to the analysis context. Therefore, overrepresentation analysis is performed to rank information items by their significance. These items originate from ontologies such as the Gene Ontology (GO) [10], which identifies cellular functions, processes and components that nodes relate to, or from KEGG [8], which relates nodes to pathways. This results in an *annotated module*, which retains the original network structure and includes enrichment information as a *set system*.

Existing tools focus on visualizing large networks, and have only limited or separate set system support or no support at all. Our proposed visual analysis approach displays sets as contours on top of a node-link layout (see Figure 2). It treats module edges and annotation sets in a unified way, and contributes the following to the analysis of annotated modules:

- Identification of elementary module analysis tasks and their composition into a visual analysis process;
- Extension of the self-organizing maps (SOM) algorithm to lay out module interactions and annotations in a unified approach;
- Implementation in the form of the Cytoscape app eXamine;
- Demonstration of eXamine via a guided study of an annotated module that is activated by the virally-encoded G protein-coupled receptor US28;

*Correspondence: m.a.westenberg@tue.nl
†Equal contributors
[1]Eindhoven University of Technology, Den Dolech 2, 5600 MB, Eindhoven, The Netherlands
Full list of author information is available at the end of the article

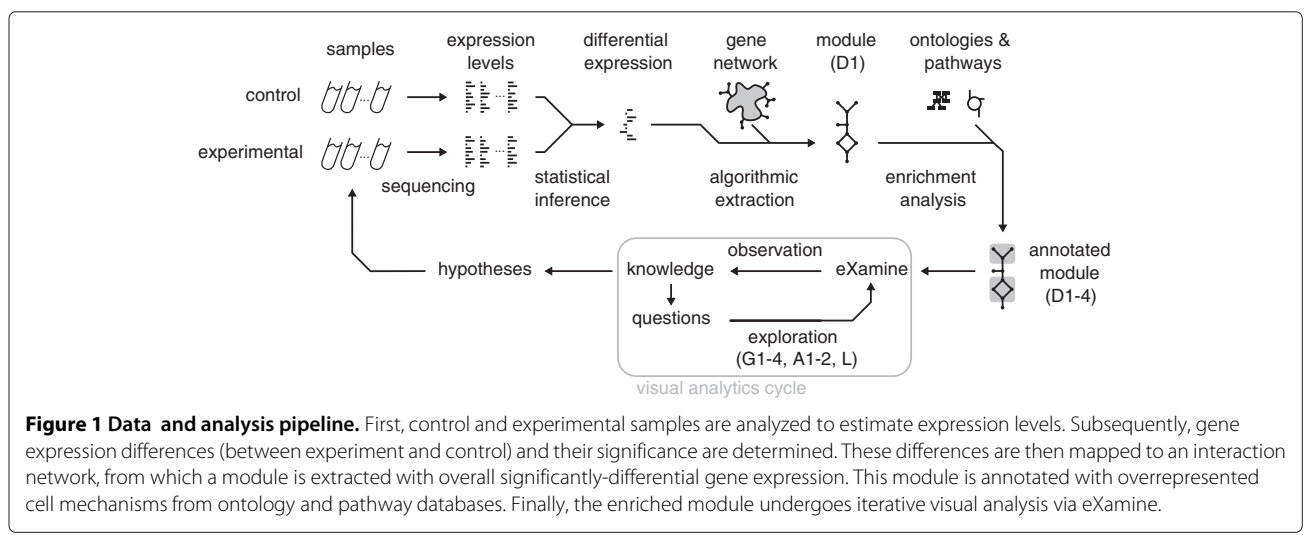

**Figure 1 Data and analysis pipeline.** First, control and experimental samples are analyzed to estimate expression levels. Subsequently, gene expression differences (between experiment and control) and their significance are determined. These differences are then mapped to an interaction network, from which a module is extracted with overall significantly-differential gene expression. This module is annotated with overrepresented cell mechanisms from ontology and pathway databases. Finally, the enriched module undergoes iterative visual analysis via eXamine.

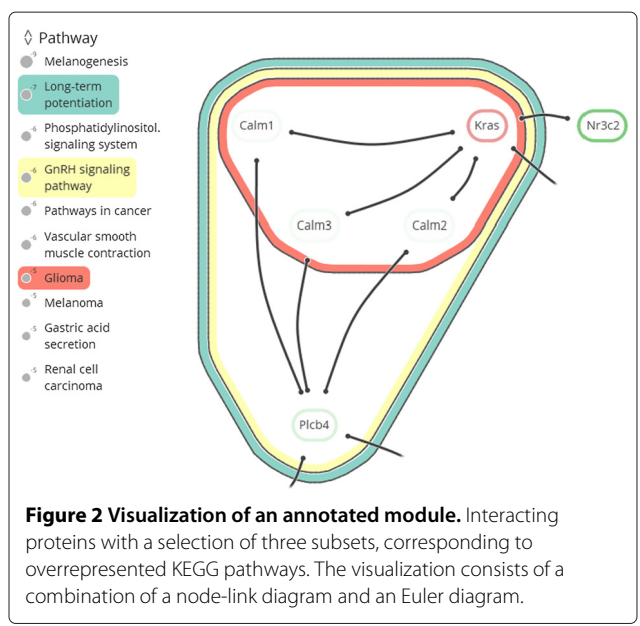

**Figure 2 Visualization of an annotated module.** Interacting proteins with a selection of three subsets, corresponding to overrepresented KEGG pathways. The visualization consists of a combination of a node-link diagram and an Euler diagram.

- Discussion on how eXamine facilitates the analysis process.

**Data characteristics**

The annotated modules—targeted by the presented method—have the following characteristics.

**D1** Small and sparse network topology, in which genes and interactions number in the dozens;
**D2** Many annotation sets, outnumbering gene interactions;
**D3** Annotation sets vary in cardinality, from a single node to the entire module;
**D4** Annotation sets overlap often.

Integrative network analysis methods produce small and sparse subnetwork modules (D1), rather than large lists of differentially expressed genes. Embedding the module in a rich context of annotations on overlapping sets of genes is a typical next step to gain insights in the underlying biology (D2, D3, D4).

**Analysis tasks**

The focus (or perspective) of analysts alternates between genes (and interactions within a module) and annotation sets. Important analysis tasks are supported for each of these data aspects to enable an analyst to hypothesize about the role of an extracted module in light of experimental conditions.

For genes, analysts want to determine:

**G1** Level of differential expression: under- or over-expressed, or insignificant;
**G2** Interacting neighbors;
**G3** Annotations (set memberships);
**G4** Annotations shared with other genes.

Single genes can become the focus of attention during the analysis process within the context of the module. The fact that a gene is part of a module does not imply that its under- or overexpression is significant. However, information (G1) about differential expression enables the elucidation of a gene's presence in the module. For example, it could be the case that a gene is not differentially expressed significantly itself, but that it is still part of a module, because it connects two differentially expressed submodules. An indirect involvement of the gene in a module mechanism is therefore likely. Neighboring genes might also become interesting (G2), as are any mechanisms that it is associated with already (G3), and the mechanisms that it shares with other genes in the module (G4).

For annotation sets, analysts want to determine:

**A1**  Significance of overrepresentation;
**A2**  Gene memberships.

If a specific gene is interesting, its annotations might be too (G3 and G4). Annotation sets themselves can have such significance (A1) that they become interesting, which then translates to genes contained in them (A2). Both significance in terms of an associated *p*-value and subjective significance are of importance to divide attention between annotation sets.

For interactions, analysts want to determine:

**L**  Annotation transitions between interacting genes.

A change between annotations (L) may occur when the focus on a gene shifts to a neighboring gene (G2), which is of importance to an analyst to judge the role and relevance of the neighboring gene in the module.

### Related work
**Network visualization and tools.** Many advanced techniques for the visualization of network topology have been developed [11-13], but few have been transferred to readily available tools. On the other hand, there are many tools for interpreting and exploring biological networks [14], including the popular open source platforms Cytoscape [15] and PathVisio [16]. However, these currently provide only limited capability to visualize annotated modules. PathVisio is a pathway analysis approach, in which sets are restricted to subsets of static, pre-defined individual pathways, and set membership is conveyed via node colors. Cytoscape's group attributes layout can be used to visualize partitions by showing disjoint parts in separate circles, but it does not support overlapping sets. The Venn and Euler diagram app [17] for Cytoscape does support overlapping sets, but it can handle only four at the same time (see Figures 3(a) and (b)). In this app, network and sets are visualized separately: set membership is conveyed by selecting a set and its corresponding nodes are highlighted in Cytoscape's network view. The RBVI collection of plugins [18] facilitates creation and editing of Cytoscape groups, and provides a group viewer that relies on aggregation of groups into meta-nodes.

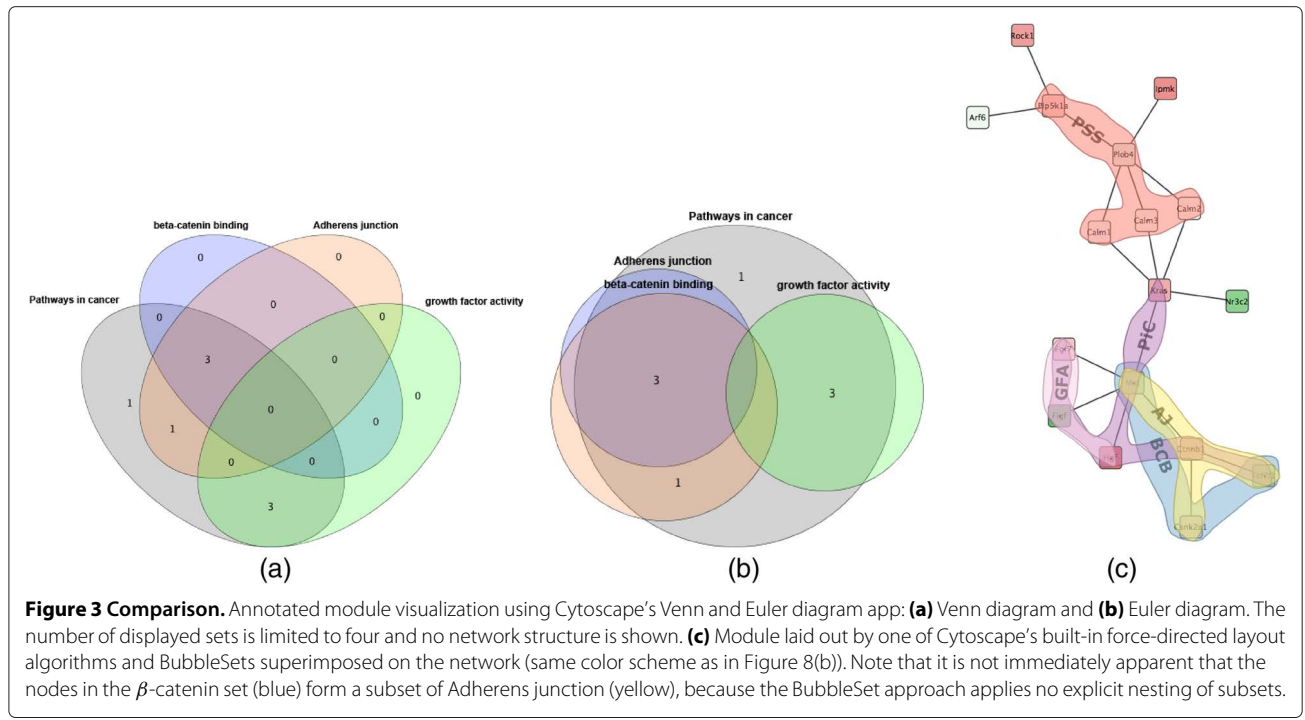

**Figure 3 Comparison.** Annotated module visualization using Cytoscape's Venn and Euler diagram app: **(a)** Venn diagram and **(b)** Euler diagram. The number of displayed sets is limited to four and no network structure is shown. **(c)** Module laid out by one of Cytoscape's built-in force-directed layout algorithms and BubbleSets superimposed on the network (same color scheme as in Figure 8(b)). Note that it is not immediately apparent that the nodes in the $\beta$-catenin set (blue) form a subset of Adherens junction (yellow), because the BubbleSet approach applies no explicit nesting of subsets.

These meta-nodes can be visualized as standard nodes, as nodes containing embedded networks, or as charts. This approach, however, does not allow for visualization of overlapping sets.

**Set system visualization.** In the information visualization field, *Euler diagrams* are used for the intuitive visualization of set systems [19-21], in which items belonging to the same set are denoted by contours. Variants of these approaches visualize sets over items with predefined positions, e.g., over a given node-link visualization of a network. These methods range from connecting these items by simple lines (LineSets) [22], via colored shapes that are routed along the items (Kelp Diagrams) [23] and contours around the items (BubbleSets, see Figure three(c)) [24,25] to hybrid approaches (KelpFusion) [26]. Visualizing an annotated module, however, requires an integrated layout of both its network and set system topologies, which is not possible with these approaches. Euler diagram methods focus on the layout of set relations at the expense of network topology. Likewise, laying out the network before superimposing set relations will emphasize network topology to the detriment of the set system. Some techniques exist that provide such integrated layouts [27-30], and which include aesthetic concerns and design of visual metaphors [31]. However, these approaches assume constraints on the network and set system topologies, e.g., strict partitions and no overlapping sets, and they are therefore not applicable to our problem.

**Method and implementation**

Visualizing an annotated module amounts to visualizing a *hypergraph* consisting of binary edges (interactions) between nodes (genes) and *n*-ary edges (annotation sets). Analysis tasks G2-G4 and A2 establish the equal importance of associating interactions and annotation sets, which reflect on both the layout as well as the visualization of the hypergraph. Therefore, as opposed to combining multiple existing techniques—e.g., a force simulation to position the nodes according to the binary edges [32], a node overlap removal algorithm to keep nodes identifiable [33], and subsequent construction of a density field to derive contours for annotation sets [24]—our approach relies on a unified algorithm that treats binary and *n*-ary edges on equal terms. This allows us to compute a balanced layout, and also to choose suitable representations for the binary and *n*-ary edges. Mathematically, we achieve this by assigning a bit vector $\mathbf{t} = (t_1, t_2, \ldots, t_M)$ to every node $t \in V$ (the module genes) that encodes its membership in binary and *n*-ary edges $S_1, S_2, \ldots, S_M$. That is, $t_i = 1$ if $t \in S_i$ and $t_i = 0$ if $t \notin S_i$.

To make this representation more concrete, consider the annotated module shown in Figure 2. The nodes are represented as the set $V = \{$*Calm1*, *Calm2*, *Calm3*, *Kras*, *Nr3c2*, *Plcb4*$\}$. There are seven sets representing the edges and three sets representing pathway memberships. The edge sets are $S_1 = \{v_1, v_4\}$, $S_2 = \{v_1, v_6\}$, $S_3 = \{v_2, v_4\}$, $S_4 = \{v_2, v_6\}$, $S_5 = \{v_3, v_4\}$, $S_6 = \{v_3, v_6\}$, and $S_7 = \{v_4, v_5\}$. Note that nodes $v_4$ (*Kras*) and $v_6$ (*Plcb4*) have some additional outgoing edges, but their targets are not visible in the image. Therefore, we ignore these edges in this example. The pathway memberships are the *Glioma* set $S_8 = \{v_1, v_2, v_3, v_4\}$, the *Long-term potentiation* set $S_9 = \{v_1, v_2, v_3, v_4, v_6\}$, and the *GnRH signaling pathway* set $S_{10} = \{v_1, v_2, v_3, v_4, v_6\}$. Now, for example, node $v_5$ gets assigned the bit vector $\mathbf{t}_{v_5} = (0, 0, 0, 0, 0, 0, 1, 0, 0, 0)$ and node $v_6$ the bit vector $\mathbf{t}_{v_6} = (0, 1, 0, 1, 0, 1, 0, 0, 1, 1)$.

This high-dimensional representation is then used to lay out the nodes without overlap, the binary edges as curves, and the *n*-ary edges as contours.

**Extension to self organizing maps**

Self Organizing Maps (*SOMs*), introduced by Kohonen [34], are artificial neural networks that are used to map high-dimensional data items to discretized low dimension. SOMs are used in a visualization setting to cluster similar items together in a 2D embedding, which results in a landscape of items based on their features [35,36]. Typical SOMs consist of a square grid of size $N \times N$ with a neuron $n_{x,y} \in [0..1]^M$ at every grid cell. A neuron $n_{x,y}$ is a bit vector of size $M$ whose dimension matches the data items' dimensions. In our case, the data items $\mathbb{T}$ correspond to the set of nodes $V$ in the annotated module. The training algorithm applies unsupervised reinforcement learning in an iterative fashion: at every iteration $i \in \{1, \ldots, I\}$ all data items $t \in \mathbb{T}$ are considered and the neuron that matches $t$ most closely is determined using a distance function such as the Euclidean or Manhattan norm. This neuron and its neighboring neurons within radius $r_i$ are updated to match $t$ even more closely by setting their respective vectors $q$ to $q + \alpha_i(t - q)$—see Figure 4(a). In early iterations *i*, the trained neighborhoods are large with $r_i$ close to the grid size $N$ and the training strength $\alpha_i$ close to 1. The parameters $r_i$ and $\alpha_i$ decrease monotonically with increasing $i$. As such, items that differ strongly will distribute across the map to establish their own regions in the grid at early stages. Items with smaller differences are separated along the grid at a more local level as the training iterations progress.

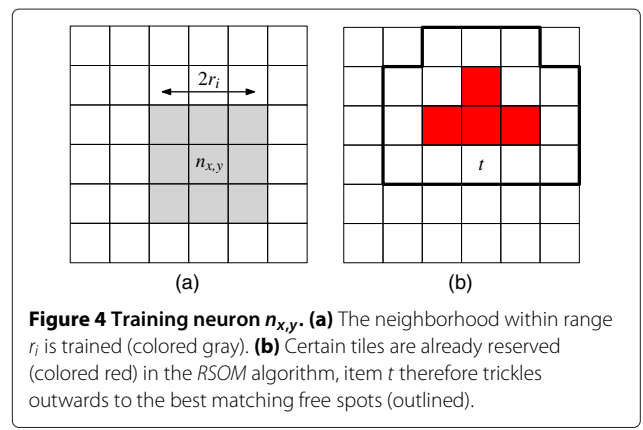

**Figure 4 Training neuron $n_{x,y}$.** **(a)** The neighborhood within range $r_i$ is trained (colored gray). **(b)** Certain tiles are already reserved (colored red) in the *RSOM* algorithm, item *t* therefore trickles outwards to the best matching free spots (outlined).

**Reservation-based training.** Similar items may end up at the same grid position in a standard SOM. This issue is usually solved by showing aggregate depictions of items, but we need to have separate depictions without overlap to support tasks G1-G4. Therefore, each item has to map to a unique grid position. We achieve this by altering the training algorithm:

**Algorithm** $RSOM(\mathbb{T})$
1.  **for** $i \leftarrow 1$ **to** $I$
2.      **do** Initialize copy $\mathbb{U}$ of $\mathbb{T}$ and clear neuron reservations.
3.          **while** $\mathbb{U}$ contains items
4.              **do** Draw and remove item $t$ from $\mathbb{U}$.
5.                  Find unreserved neuron $n_{x,y}$ with smallest distance $d(t, n_{x,y})$.
6.                  Reserve $n_{x,y}$ for $t$.
7.                  **for** any neuron $q$ within range $r_i$ from $(x, y)$
8.                        **do** $q \leftarrow q + \alpha_i (t - q)$

The algorithm assigns items to a unique neuron after every training iteration, because, once a neuron is reserved by an item, subsequent items will ignore it. This causes a flooding effect where similar items end up in the same area of the grid and trickle outwards as the area becomes more crowded—see Figure 4(b).

**Configuration.** The metric distance form of cosine similarity is used as the distance function $d$, i.e. $d(q, p) = \cos^{-1}((q \cdot p)(|q||p|))\pi^{-1}$. This measure outperforms the Euclidean and Manhattan norms in high-dimensional spaces. The SOM is trained with a learning strength and neighborhood range that decrease linearly with increasing iteration $i$. A standard choice is $\alpha_i = c \cdot (1 - i/I)$ and $r_i = \lfloor (1 - i/I) \cdot N \rfloor$, where $c \in (0..1)$ is a small constant that determines the initial training strength. We use $N = 2|\mathbb{T}|$ for the number of neurons and iterations, balancing node placement freedom versus required display space, and $I = 10^6/|\mathbb{T}|$ for a gradual and accurate training, respectively.

**Layout preservation.** A new layout has to be computed whenever the user selects or deselects a set. The new layout should change little in comparison to the old layout to preserve the user's mental map. This is achieved by a simple addition to the SOM algorithm, where a new SOM is initialized with the previous configuration of the neurons, i.e., an item that was positioned at $n_{x,y}$ in the old SOM is placed at $n_{x,y}$ in the new SOM and its neighborhood is trained according to the new bit vector of the item. The new SOM retains much of the initial configuration by starting the training factor $\alpha_i$ at $c = 0.01$. Naturally, this imposes a trade-off between layout quality and conservation. The layout will sometimes change strongly to accommodate the addition of a set that contains many items. In contrast, the layout can be retained if only a small set that does not alter much of the topology is added. This approach does not consider a history of topological changes, as is done in online graph drawing [37] to capture temporal dynamics, but is sufficient to maintain a stable and interactive environment.

**Set dominance.** The user is enabled to make a certain set more dominant in the layout by having the training algorithm place the items of that set closer to each other than the items of other sets. This relies on weighting the components of the item bit vectors: every $S_i$ is given a weight $w_i$ with $w_i = 1$ initially. The bit vectors are augmented to incorporate these weights: $t_i = w_i$ if $t \in S_i$ and $t_i = 0$ if $t \notin S_i$. The bit vector component of $S_i$ will therefore play a more prominent role in distance metric $d$ when the user increases $w_i$—see Figure 5.

Assigning greater weight to a set improves the quality of its layout by coalescing its elements, which aids tasks G4 and A2. However, it also degrades the layout quality of other sets and links when their topology conflicts with the prioritized set. This stems from the difficulty of projecting elements from a high-dimensional space down to a two-dimensional space, which sometimes results in a suboptimal layout per set. Interactive manipulation provides a way to assign different priorities to sets, and improve their layouts.

**Contours.** The SOM's neuron grid is used to define the contours representing the active set system. Let $S_i$ be an active set. The corresponding $i$-th components of the neurons define a scalar field that forms a fuzzy membership landscape for $S_i$. This field is similar to the density field used in Bubble Sets [24]. Now, the inclusion of the grid tile of neuron $n$ in the contour body is determined by imposing a threshold, of for example $\frac{1}{2}$, on the $i$-th component (see Figure 6(a)). The contour can then be tightened to reduce sharp corners by including parts of tiles that are free of items, as illustrated in Figure 6(b).

After establishing the layout of the contours, we apply geometric post processing steps [23] to improve aesthetics, where all sets are legible (tasks G3 and A2) and contours form clear boundaries underneath interactions (task L). Sharp corners of the initial contours are rounded by a dilation of *r*, erosion of 2*r*, and subsequent dilation of *r* (see Figure 6). Here *dilate* and *erode* are equivalent to *Minkowski sum* and *Minkowski subtraction* operators with a circle of radius *r* [38], respectively. In addition, the contours are nested by applying different levels of erosion, enforcing a certain distance between them. The thick colored ribbons in Figure 7 are obtained by taking the body *b* of a contour, eroding it to get a smaller body $b_e$, and taking the symmetric difference $b - b_e$ of *b* and $b_e$ to effectively cut $b_e$ out of *b*. Here, the extent of the erosions and dilations (radius *r*) is bounded by a fraction of the grid's tile size. This guarantees that items are contained by a contour of $S_i$ if, and only if, these items are contained by $S_i$.

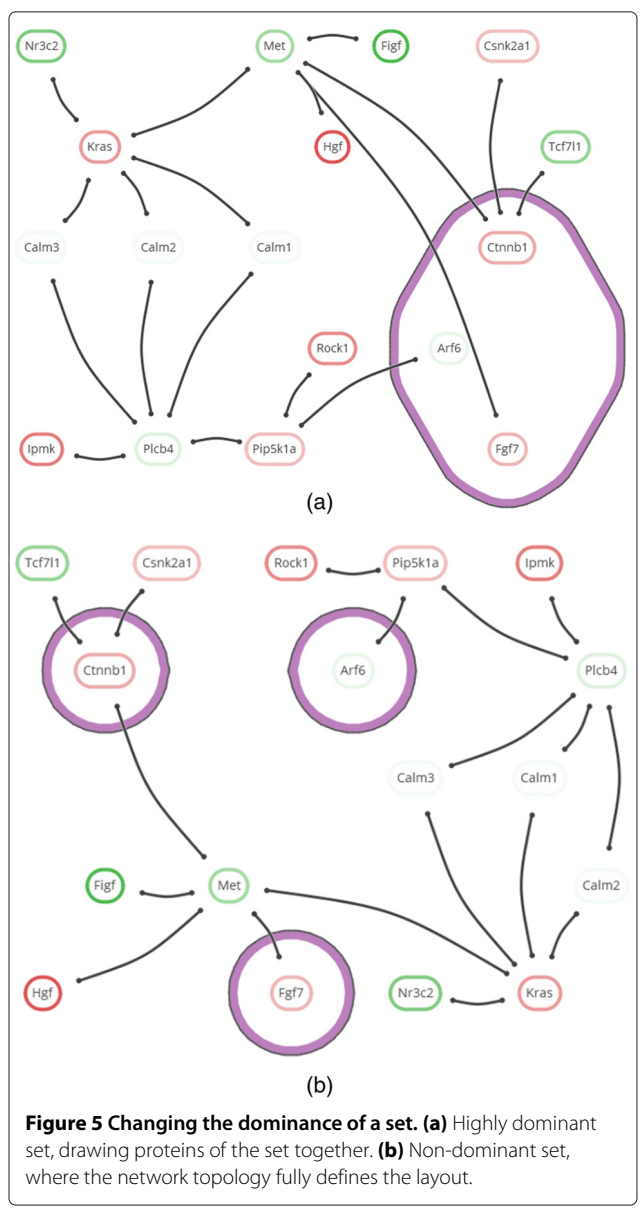

**Figure 5 Changing the dominance of a set. (a)** Highly dominant set, drawing proteins of the set together. **(b)** Non-dominant set, where the network topology fully defines the layout.

Set contours are drawn in descending nesting order, which is defined by their different erosion levels; the largest contour is drawn first and the smallest contour last. The contour ribbons are assigned unique colors per set and are drawn fully opaque to prevent any confusion caused by blended colors. Occlusion is mitigated by limiting the width of the ribbons. Finally, the contours are drawn a second time as dashed lines such that occluded contour sections can be inferred—see Figure 7.

**Implementation**

We have implemented the technique in a Cytoscape app, and have emphasized simplicity of interaction and visual presentation in the design. The available sets are sorted by significance and listed in the *set overview* on the left, where the significance of a set is visualized as a circle, scaled logarithmically and accompanied by its scientific exponent as text (task A1). The user may select sets for inclusion in the annotated *network visualization* to the right—see Figure 8(c). All described functionalities can be used at interactive speeds for networks up to dozens of nodes, edges, and active sets, including laying out the network with the *RSOM* training algorithm. Geometric operations on the contours, such as dilations and erosions, are performed via Java Topology Suite [39].

**Interaction.** Interactions consist of simple mouse actions (see the video in the Supplemental Material). The inclusion of a set in the network visualization is toggled via the set's label in the set overview or its contour in the network visualization (task A2). Additional information about a set or node may be obtained via a hyperlink to a web page provided in the input data, enabling quick access to external information sources such as the KEGG website. This approach keeps the tool flexible, i.e., the tool itself does not have to be altered every time a new kind of set or node from a different database is loaded.

The links of a node are emphasized when it is hovered over (see Figure 9(a)) such that its direct neighborhood can be discerned from its surroundings (task G2). Moreover, sets that contain the hovered node are highlighted as well. Likewise, links can be hovered to highlight their nodes and common sets. Vice versa, the contours of a set are emphasized and its comprising nodes are highlighted when it is hovered over (see Figure 9(b)). This provides immediate feedback to the user about node-set relations (tasks G3 and A2) without having to select a set and consequently changing the layout of the network visualization.

The lists of annotations sets can be expanded and collapsed by clicking on their headers, and scrolled downward to sets of lower significance by turning the mouse wheel. The set circles that convey significance remain visible at all times, grouping at the list top and bottom, to guarantee the depiction of all set memberships when a node is hovered.

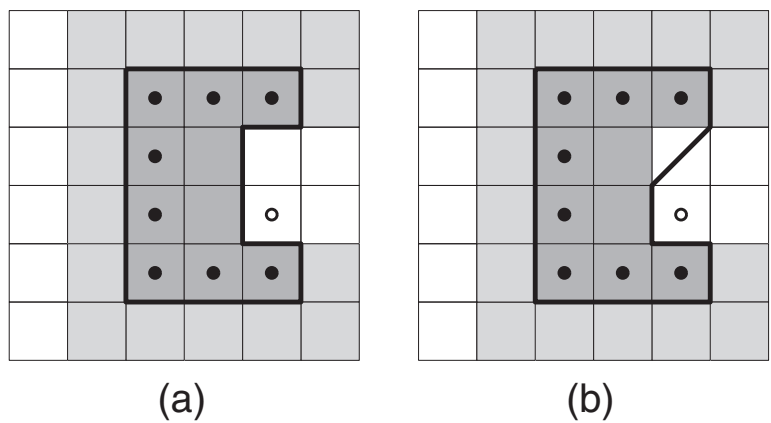

**Figure 6 Derivation of contours for set $S_i$.** The darkness of a tile represents the value of the neurons' *i*-th component, the thick black line is the contour, dots represent items that are in $S_i$, and white dots are items that are not in $S_i$. **(a)** Contour that results from the union of tiles with a value above a certain threshold. **(b)** Refined contour with shortcuts across free tiles.

The user can adjust the dominance of a set by spinning the mouse wheel while hovering over either the set's label in the set overview or contour in the network visualization. This enables the user to give a set a central role in the layout (see Figure 5(a)) or to remove any of its influence (see Figure 5(b)).

All changes to the visualization caused by interaction are animated. Colors and positions of items are altered gradually. Link layout changes are animated by interpolating their control points, while contour layouts are handled by fading out the old contour and fading in the new contour. The use of layout preservation, as described previously, in combination with animations helps to preserve the user's mental map.

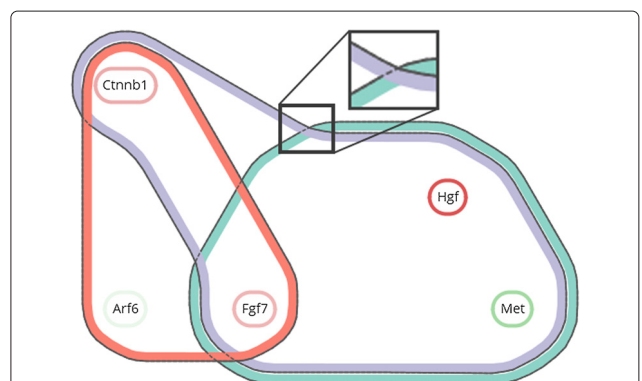

**Figure 7 Geometric refinement of set contours after initial layout.** Corners are smoothened by dilation and erosion operations, and contours are given a thick and colored internal ribbon. Unique erosion levels create distance between contour outlines, and contour overlap is emphasized by dashed lines.

**Color.** Unique, distinguishable colors are derived from Color Brewer palettes [40], and assigned to annotation sets in a cyclic manner to avoid assigning the same color consecutively. In addition, large differences in contrast are avoided. For example, text and set outlines are colored dark gray instead of black to reduce their visual dominance. Black is only used when items are hovered over or highlighted such that they attract attention, as shown in Figure 9. Moreover, labels of selected sets (in the set overview) are emphasized with a more intense black color to ensure that they are readable in a colored surrounding. Node labels have a white background to make sure that their text is legible when drawn on top of a set ribbon with a dark color. Likewise, links have halos that make them easier to distinguish and their intersections more pronounced.

**Cytoscape integration.** eXamine is tightly integrated into Cytoscape. Cytoscape's group functionality is used to represent sets and we rely on the table import functionality for importing both the set and node annotations. The user is also able to group sets into different categories. The Cytoscape node fill color map attribute is used to color the nodes in eXamine according to gene expression score (task G1). The user therefore has the freedom to define the desired color map via Cytoscape. The user can invoke eXamine on the currently selected nodes via the eXamine control panel. There the user can select which categories to show as well as the number of sets per category. In addition, the user can specify that the Cytoscape selection should be updated to match the union or intersection of the selected sets in eXamine. This enables the use of eXamine with any kind of module extraction algorithm and/or filter method in Cytoscape, which includes manual node selection.

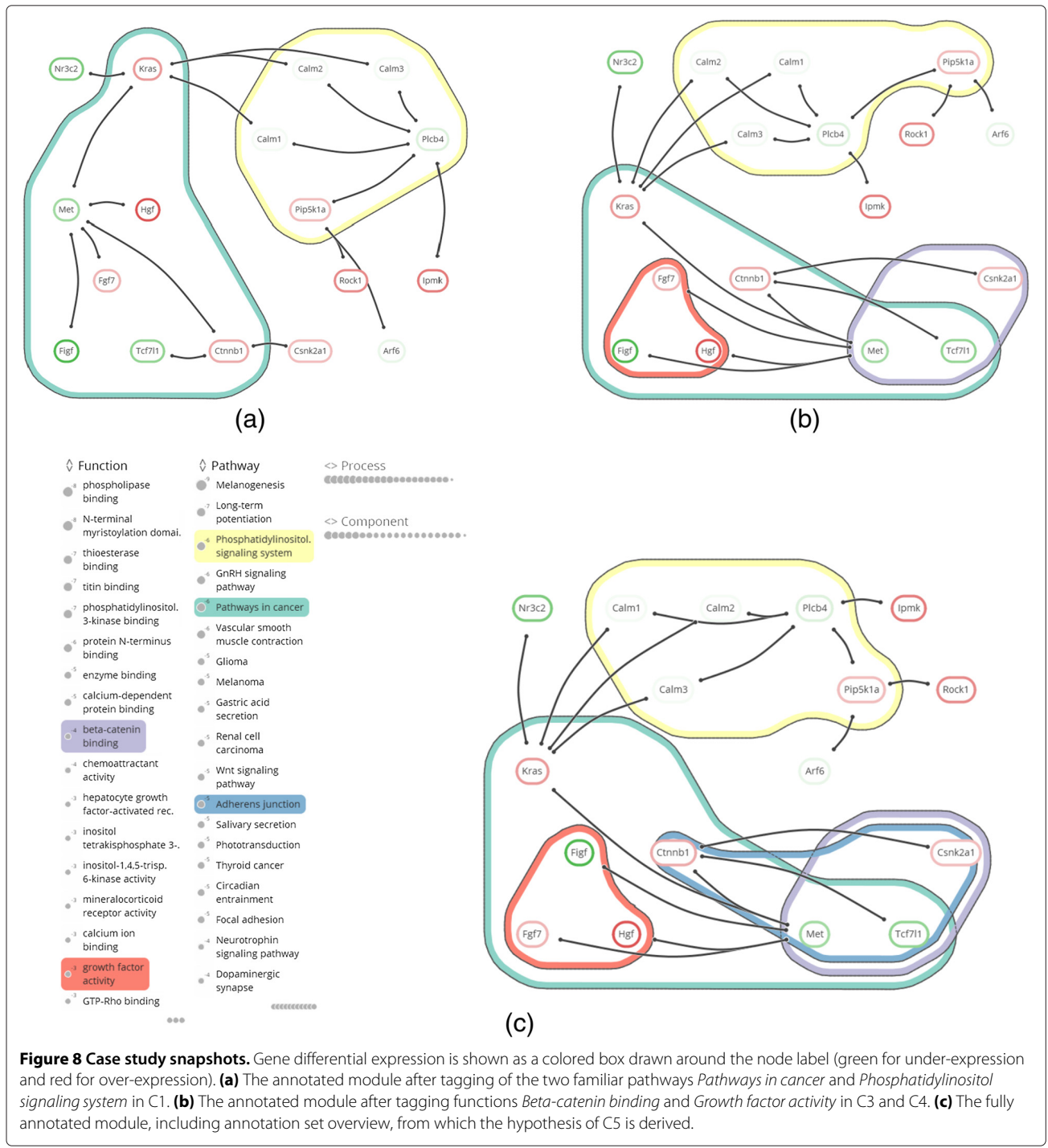

**Figure 8 Case study snapshots.** Gene differential expression is shown as a colored box drawn around the node label (green for under-expression and red for over-expression). **(a)** The annotated module after tagging of the two familiar pathways *Pathways in cancer* and *Phosphatidylinositol signaling system* in C1. **(b)** The annotated module after tagging functions *Beta-catenin binding* and *Growth factor activity* in C3 and C4. **(c)** The fully annotated module, including annotation set overview, from which the hypothesis of C5 is derived.

### Results: a case study of US28 mediated signaling

We demonstrate how a domain expert can use eXamine by working out a case study in which a data set is re-analyzed (this work was done by the co-authors with biological expertise). While this data set has been studied extensively, it was possible to derive a new hypothesis via eXamine

The *Human Cytomegalovirus (HCMV)* is a highly-contagious herpes virus [41]. Infection with HCMV in healthy humans usually does not result in symptoms. However, in humans with a compromised immune system the virus is correlated with diseases such as hepatitis and retinitis [42]. In addition, HCMV gene products have been detected in various tumors even though HCMV is not considered to be an oncogenic virus. Experts therefore hypothesize that the virus may act as a stimulating factor during onset and development of cancer without being a root cause [43-45].

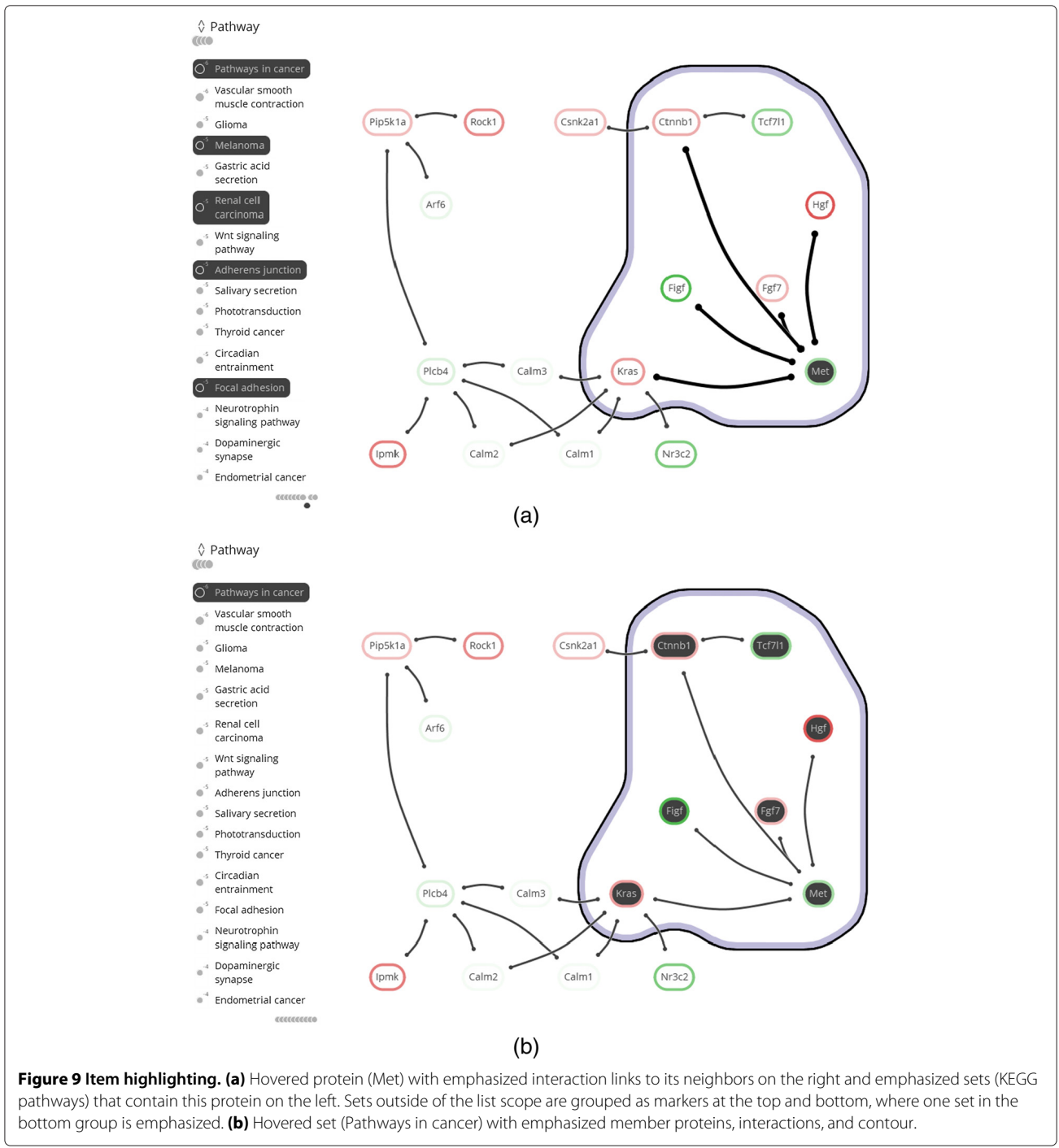

**Figure 9 Item highlighting. (a)** Hovered protein (Met) with emphasized interaction links to its neighbors on the right and emphasized sets (KEGG pathways) that contain this protein on the left. Sets outside of the list scope are grouped as markers at the top and bottom, where one set in the bottom group is emphasized. **(b)** Hovered set (Pathways in cancer) with emphasized member proteins, interactions, and contour.

HCMV is responsible for the production of several viral G protein-coupled receptors (vGPCRs). Of these vGPCRs, US28 is the most studied and is characterized as chemokine sink [46]. Chemokines are signaling proteins that induce cell migration. Moreover, US28 hijacks the host cell's signaling pathways, stimulates proliferative signaling pathways [47-51]. Previous studies focused on transcriptome analysis to evaluate pathways that are affected by US28. Differentially expressed genes involved in HCMV-induced disease symptoms were identified and related to known pathways [49,50]. However, this analysis did not include network-based module extraction and enrichment.

To identify additional deregulated signaling due to US28, we analyzed the same data overlaid on the KEGG mouse network [8]. The network consisted of 3863 nodes and 29293 edges. Gene *p*-values, reflecting whether genes are significantly differentially expressed, were derived using RMA [52] and LIMMA [53]. Heinz [7], a tool for identifying differentially expressed modules, was then applied using a false discovery rate of 0.0007. This resulted in a module of 17 proteins. Finally, enrichment analysis using TopGO [54] was performed to annotate this module with enriched GO-terms and KEGG pathways (see Figure 8).

These data processing steps correspond to the initial steps in Figure 1. The subsequent analysis of the annotated module aims at obtaining new insights about US28-mediated signaling. The analysis follows the visual analytics cycle consisting of *observation*, *knowledge*, *questions* and *exploration*, finalized by a hypothesis. All analysis steps are shown in the screencast of Additional file 1.

### C1 Two familiar pathways
**Observation.** The KEGG pathway annotation sets show significant presence of *Pathways in cancer* and *Phosphatidylinositol signaling* (*p*-values of $5.6 \cdot 10^{-6}$ and $1.0 \cdot 10^{-6}$, respectively).

**Knowledge.** An oncomodulatory role has been proposed for US28 [43-45], which coincides with the presence of *Pathways in cancer* and makes the genes annotated by this term of interest. *Phosphatidylinositol signaling* corresponds to previous work linking US28 to Phosphatidylinositol-mediated calcium responses [47,55].

**Question.** Which parts of the module are involved in *Pathways in cancer* and *Phosphatidylinositol signaling*?

**Interaction.** Tag the *Pathways in cancer* and *Phosphatidylinositol signaling* annotation sets (see Figure 8(a)).

### C2 Choosing sides
**Observation.** Clear division of the module is apparent after tagging the two familiar pathways. Genes *Arf6*, *Csnk2a1*, *Csnk2a1*, *Ipmk*, *Nr3c2* and *Rock1* are not part of the pathways but have direct, unambiguous interactions with either of the pathways.

**Knowledge.** Because of the known involvement of US28 in *Phosphatidylinositol signaling*, we do not focus on the genes of this pathway (*Calm1..3*, *Plcb4*, *Pip5k1a*), nor on the directly interacting genes (*Arf6*, *Ipmk*, *Rock1*). Instead, the *Pathways in cancer* genes *Kras*, *Met*, *Figf*, *Hgf*, *Fgf7*, *Ctnnb1* and *Tcf7l1*, and directly interacting genes *Nr3c2* and *Csnk2a1* may lead to new insights in US28-mediated signaling and ultimately the oncomodulatory role of HCMV.

**Question.** Do any of the aforementioned genes in or adjacent to *Pathways in cancer* lead to new insights in US28-mediated signaling?

**Interaction.** Hover over the genes in and close to *Pathways in cancer* to determine mechanisms of interest.

### C3 A twist of β-catenin
**Observation.** The genes in *Pathways in cancer* can be divided roughly into two subsets: those that are annotated by *growth-factor activity* and those annotated by *β-catenin binding* (see Figure 8(b)). *Csnk2a1*, *Tcf7l1* and *Met* are part of the latter annotation set, where *Tcf7l1* and *Csnk2a1* are down- and up-regulated, respectively. Expression of the neighboring *Ctnnb1* (*β*-catenin) is up-regulated.

**Knowledge.** *β*-catenin signaling results in elevated protein levels of the TCF/LEF transcription factor family that contains the protein encoded by *Tcf7l1*. Although *Tcf7l1* is down-regulated, a recent study shows that this is not reflected at the protein level and that US28 induces *β*-catenin signaling [51]. In the same study, involvement of WNT/Frizzled via the canonical signaling pathway was ruled out and a hypothesis stating that US28-mediated signaling of *β*-catenin proceeds via ROCK1, which is also present in the module, was postulated.

**Question.** Are there alternative mechanisms explaining the activation of *β*-catenin?

**Interaction.** Tag the *Growth factor activity* annotation set (see Figure 8(b)).

### C4 Growing knowledge
**Observation.** *Fgf*, *Hgf* and *Figf* are annotated with *Growth factor activity* and connected to *β*-catenin via *Met*.

**Knowledge.** MET is a receptor tyrosine kinase, whose only ligand is HGF. Therefore we can rule out the links from *Met* to *Fgf* and to *Figf*. In fact, these links are artifacts of how the mouse network was constructed from KEGG pathways. These artifacts often link whole groups of genes such as, in this case, growth factors to receptor tyrosine kinases.

**Question.** Does the *Hgf*–*Met* axis relate to *β*-catenin activation?

**Interaction.** Hover over *Met* and *Ctnnb1* (*β*-catenin).

### C5 New insights
**Observation.** *Met* and *β*-catenin are both part of the *Adherens junction* pathway, as are *Tcf7l1* and *Csnk2a1* (see Figure 8(c)).

**Knowledge.** Adherens junctions bind two cells together, keeping multiple cells in place. Alternative mechanisms have been described that explain $\beta$-catenin activation via the release of $\beta$-catenin from cell to cell adherens junctions (e.g. [56]). US28 promotes cell migration [57,58], which causes the loss of cell to cell contacts with subsequent release of $\beta$-catenin into the cytoplasm. This may explain increased levels of $\beta$-catenin as found previously [51].

By requesting additional information for *Adherens junction* via eXamine, showing an external website by KEGG, we find an indirect connection between *Met* and $\beta$-catenin in the pathway (see Figure 10). Activation of MET via HGF mediates the release of $\beta$-catenin from adherens junctions, resulting in increased TCF/LEF levels [59,60].

**Hypothesis.** Combining this with the growth factor observations of C4 leads to the following hypothesis.

- US28-mediated up-regulation of *Hgf* results in elevated levels of the corresponding HGF protein;
- The subsequent activation of MET results in the release of $\beta$-catenin into the cytoplasm;
- Subsequent translocation into the nucleus leads to enhanced TCF/LEF activation.

*Synopsis*

We are currently validating the hypothesis experimentally. Preliminary results indicate that the up-regulation of *Hgf* is indeed reflected at the protein level. Should this hypothesis turn out to be true, we would obtain crucial insights into one of the mechanisms by which the HCMV-encoded chemokine receptor US28 rewires cellular signaling. Ultimately, we would like to understand how this virus achieves its oncomodulatory role and how this can be disrupted.

**Discussion**

The analysis tasks described in the background section guided the design decisions that we have taken in the implementation of eXamine. These decisions are motivated via the analysis cycles of the US28 case study.

**Overview.** The benefit of a spacious annotation set overview follows from the first cycle (C1), in which the categorized, ranked, and legible annotation lists enable the fast recognition of two familiar and significantly represented pathways (task A1). Subsequent tagging of the two pathways reveals their module genes (task A2) and concisely drawn contours emphasize the division of the module into two parts and some additional genes that are not part of the pathways.

An annotation table, separate of the network, would not have made this division as apparent. The main reason is that annotation set transitions along gene interactions are not explicit in such a representation. In contrast, such cross-contour interaction links are clearly visible in eXamine (e.g. the transition from *Kras* in *Pathways in cancer* to *Nr3c2* outside of *Pathways in cancer*).

**Annotated genes.** The need to focus on specific genes and their properties appears in the second analysis cycle (C2), in which genes of *Pathways in cancer* are inspected for annotations of interest (task G3). Highlighting annotations by hovering over genes enables fast identification of relevant annotations in the stable overview that oriented the analyst in C1. Vice versa, hovering an annotation of interest ($\beta$-*Catenin binding*) confirms that it is shared by *Csnk2a1*, *Tcf7l1*, and *Met* (task G4). The same observations could have been made from an annotation table. However, the topological characteristics of these three genes would have been harder to discern, i.e., their direct interaction with *Ctnnb1* (task G2). This also applies to other set visualizations without depiction of network topology, such as Venn or Euler diagrams, as shown in Figure 3(a) and (b). To make the topology of the gene interactions more explicit, a node-link visualization could be used. For example, Figure 3(c) shows the module laid out by one of the built-in force-directed layout algorithms of Cytoscape with all five annotation sets superimposed as BubbleSets. However, the structure of the annotation sets is hard to discern, and it is not immediately clear that nodes belonging to the $\beta$-catenin binding set (blue shape) form a proper subset of the Adherens junction set (yellow shape).

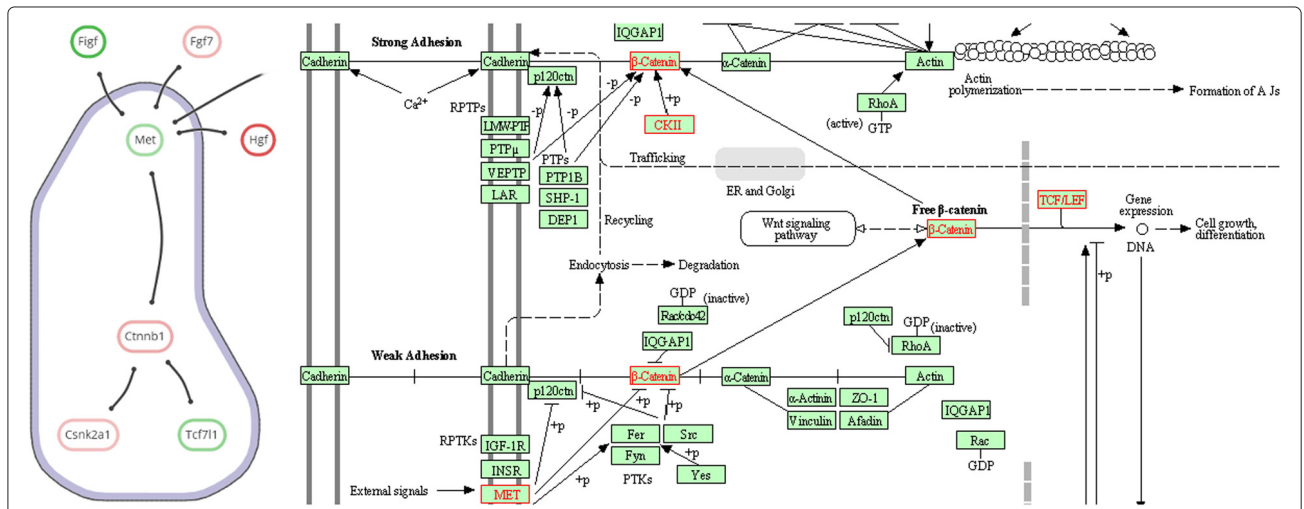

**Figure 10 Connection between Met and $\beta$-catenin.** Proteins that are associated to the selected *Adherens junction* at the left and corresponding KEGG pathway information at the right, where reactions catalyzed by module proteins are marked in red. Activation of MET by its ligand HGF results in the phosphorylation of $\beta$-catenin. This in turn results in its release from cadherin-complexes on the cell membrane into the cytoplasm.

**Integration.** The third cycle (C3) shows the importance of gene expression values (task G1), which is not limited to the interpretation of genes in isolation but along multiple genes, their interactions, and shared annotation sets. The importance of integrated support for all analysis tasks follows from the remaining cycles (C4-C5), where multiple deductions are made in succession via multiple tasks. Here, tagging relevant pathways enables the analyst to build up a context for making deductions.

**Limitations.** eXamine is designed to accurately convey small and annotated modules, consisting of up to about thirty proteins and categories of up to about twenty annotations (note that these limits are not hard). The case study shows that common analysis tasks for these modules are covered. Scalability is a concern as our approach focuses on small modules to enable accurate depiction of sets contours; it is not possible to construct a comprehensive layout if the module consists of hundreds of proteins or if there are dozens of annotation sets to visualize at the same time. Both aspects make visual analysis ineffective. This is a natural limitation of any visualization approach based on node-link diagrams and set contours, however.

Our technique relies on a focus and context approach, in which the network and set system has been pruned down to the most relevant components first. Communicating small-scale information is given priority to support hypothesis generation at the level of individual proteins and their interactions, as follows from the targeted analysis tasks. Nonetheless, the tool is capable of visualizing modules of up to a hundred proteins, albeit with less legibility of interactions and annotations.

The integration of eXamine into Cytoscape mitigates many scalability issues. Cytoscape, for example, provides a global view of the network, in which the user can zoom in on smaller subnetworks for more in-depth analysis by eXamine. In addition, the integration into Cytoscape provides access to further analysis algorithms.

The extended SOM algorithm embeds an annotated module to reflect its topology, i.e., the distances between its proteins based on common interactions and annotations. This does not guarantee optimal aesthetics however, and unnecessary link and contour intersections can sometimes occur. The analysis tasks targeted by eXamine are not much hampered by such intersections since all interactions, annotations, and their interplay remain pronounced. However, to communicate analysis results, aesthetics might need further improvement. This could be done by weighing aesthetic criteria such as the number of intersections and shape complexity against each other, and formulating this as a combinatorial optimization problem. The associated algorithms [11] are often complex, and it is not so easy to integrate them into an interactive system.

**Application to other domains.** eXamine is not limited to the analysis of enriched protein modules nor to data from the biological domain. It can be applied to any small network module that is accompanied by a set system, such as a social circle that consists of people, their relationships, and common interests.

## Conclusions

We have proposed a visualization approach that enables the analysis of small and annotated network modules, and have implemented this in the Cytoscape app eXamine. Our approach displays sets as contours on top of a node-link layout. We have introduced an extension to the self-organizing maps algorithm to lay out module edges and annotation sets in a unified way. The added value of our approach has been demonstrated in a case study of a US28-mediated signaling module, in which a novel hypothesis about the way US28 induces $\beta$-catenin signaling has been derived.

## Availability and requirements

**Project name:** eXamine
**Project homepage:** http://apps.cytoscape.org/apps/examine
**Operating system(s):** all
**Programming language:** Java
**Other requirements:** Cytoscape 3.x
**License:** GPL2
**Any restrictions to use by non-academics:** None

## Additional file

**Additional file 1: Screencast.** Screencast of interactive analysis in eXamine for the US28 case study.

**Competing interests**  
The authors declare that they have no competing interests.

**Authors' contributions**  
KD, MEK, MAW and GWK conceived the visual analysis technique. KD and MEK implemented eXamine. CIB, MS and MJS applied it to the US28 case study after instructions by MEK and GWK. KD, MEK, MAW and GWK drafted the manuscript. All authors read and approved the final manuscript.

**Acknowledgements**  
Kasper Dinkla is supported by the Netherlands Organisation for Scientific Research (NWO) under project no. 612.001.004.

**Author details**  
[1]Eindhoven University of Technology, Den Dolech 2, 5600 MB, Eindhoven, The Netherlands. [2]Life Sciences, Centrum Wiskunde & Informatica (CWI), Science Park 123, 1098 XG, Amsterdam, The Netherlands. [3]VU University Amsterdam, De Boelelaan 1105, 1081 HV, Amsterdam, The Netherlands.